\documentstyle[12pt,fleqn,epsfig]{article}
\newcommand{\be}{\begin{eqnarray}}
\newcommand{\ee}{\end{eqnarray}}
\newcommand{\ba}{\begin{array}}
\newcommand{\ea}{\end{array}}
\newcommand{\half}{{\textstyle{\frac{1}{2}}}}
\newcommand{\partialslash}{\partial\hspace{-.5em}/\hspace{.15em}}
\newcommand{\Pslash}{P\hspace{-.5em}/\hspace{.15em}}

\newcommand{\kslash}{k\hspace{-.5em}/\hspace{.15em}}

\newcommand{\kint}{\int_{\Lambda_3}\!\frac{d^4 k}{(2\pi)^4}}

\begin{document}
%

\begin{flushright}
HUB-EP 97/28 \\ hep-th/9705334 \\ May 1997
\end{flushright}

\begin{center}
{\Large\bf  
Excited pions, $\rho$- and $\omega$- mesons and their decays  \\
in a chiral $SU(2)\times SU(2)$ Lagrangian } \\[1.5cm]
%
{\large\bf M.K.\ Volkov}$^{\rm 1}$ \\[0.4cm]
{\em Bogoliubov Laboratory of Theoretical Physics \\
Joint Institute for Nuclear Research \\
141980, Dubna, Moscow region, Russia}
\\[0.7cm]
\vskip0.5truecm
{\large\bf D.\ Ebert}$^{\rm 2}$ \\[0.4cm]
{\em Institut f\"ur Physik \\
Humboldt-Universit\"at zu Berlin \\
Invalidenstrasse 110, D-10115 Berlin, Germany}
\\[0.7cm]
\vskip0.5truecm
{\large\bf M.\ Nagy}$^{\rm 3}$ \\[0.4cm]
{\em Institute of Physics \\
Slovak Academy of Sciences \\
842 28 Bratislava, Slovakia}
\\[0.7cm]
\end{center}
\vspace{0.5cm}
\begin{abstract}
\noindent
A chiral $SU(2)\times SU(2)$ Lagrangian containing,
besides the usual meson fields, their first radial excitations
is considered. The Lagrangian is derived by bosonization of the
Nambu--Jona-Lasinio quark model with separable non-local
interactions, with form factors corresponding to 3--dimensional
ground and excited state wave functions. The spontaneous breaking
of chiral symmetry is governed by the NJL gap equation. The first
radial excitations of the pions, $\rho$-and $\omega$-mesons are described
with the help of two form factors. The weak decay constant $F_{\pi'}$
is calculated. The values for the decay widths of the processes
$\rho \to 2\pi$, $\pi' \to \rho\pi$, $\rho'\to 2\pi, 
\rho' \to \omega \pi$ and $\omega' \to \rho \pi$ are obtained
in agreement with the experimental data. 
\end{abstract}
%
%
\rule{5cm}{.15mm}
\\
\noindent
{\footnotesize $^{\rm 1}$ E-mail: volkov@thsun1.jinr.dubna.su} \\
\noindent
{\footnotesize $^{\rm 2}$ E-mail: debert@qft2.physik.hu-berlin.de} \\
\noindent
{\footnotesize $^{\rm 3}$ E-mail: fyzinami@nic.savba.sk} \\
\newpage

\section{Introduction}
So far the experimental status of excited light mesons like the $\pi'$
and $K'$ is not yet completely established, requiring further investigations
both in experiment and theory \cite{Rev_96,volk_96}. In particular,
the theoretical study of radially (orbitally) excited mesons 
is expected to provide us with a deeper understanding of the internal
structure of hadrons and, equivalently, of the underlying effective
interquark forces.
\par
 In the previous papers \cite{volk_96,volk_97} of one of the authors (MKV) 
a simple extension of the NJL-model with nonlocal separable quark 
interactions for the description of radially excited mesons was proposed. 
The theoretical foundations for the choice of the polynomial pion-quark form 
factors were discussed and it was shown that we can choose these form 
factors in such a way that the mass gap equation conserves the usual form 
and gives a solution with a constant constituent quark mass. Moreover, the 
quark condensate does not change after including the excited states in the 
model, because the tadepoles connected with the excited scalar fields vanish.
Thus, in this approach it is possible to describe radially excited mesons
above the usual NJL vacuum preserving the usual mechanism of 
chiral symmetry breaking. Finally, it has been shown that one can derive an 
effective meson Lagrangian for the ground and excited meson states directly 
in terms of local fields and their derivatives. The nonlocal separable 
interaction is defined in Minkowski space in a 3-dimensional (yet 
covariant) way whereby form factors depend only on the part of the quark-
antiquark relative momentum transverse to the meson momentum.
This ensures the absence of spurious relative-time excitations 
\cite{feynman_71}. 
\par
In paper \cite{volk_97} the meson mass spectrum for the ground and excited 
pions, kaons and
the vector meson nonet in the $U(3) \times U(3)$ model of this 
type was obtained. By fitting the meson mass spectrum all parameters in this model are fixed. This then allows one to describe all the strong, 
electromagnetic and weak interactions of these mesons without introducing 
any new additional parameters. 
\par
In the papers \cite{volk_96,volk_97} the weak decay constants $F_{\pi'}, 
F_{K}$ and $F_{K'}$ were described. In the present work we would like to
extend 
this by demonstrating that this model satisfactorily describes two types of
decays. This concerns 
strong decays like $\rho \to 2 \pi, \pi' \to \rho \pi$, $\rho' \to 2\pi$
associated to divergent quark diagrams as well as the decays
$\rho' \to \omega \pi$ and $\omega' \to \rho \pi$ defined by anomalous quark
diagrams.

\par
The paper is organized as follows.
In section 2, we introduce the effective quark interaction
in the separable approximation and describe its bosonization.
We discuss the choice of form factors necessary to describe the
excited states of the scalar meson, pions, $\rho$, $\omega$ and $a_1$-mesons.
In section 3, we derive the effective Lagrangian for the pions
and perform the diagonalization leading to the physical pion
ground and excited states. In section 4, we perform the diagonalization
for the 
$\rho$ and $\omega$-mesons. In section 5, we fix the parameters of our model
and evaluate the masses of the ground and excited states of pions and
$\rho$-mesons and the weak decay constants $F_{\pi}$ and $F_{\pi'}$.
In section 6, we evaluate the decay widths of the processes
$\rho \to 2 \pi$, $\pi'\to \rho \pi, \rho' \to 2\pi, \rho' \to \omega \pi$ 
and $\omega' \to \rho \pi$. The obtained results are
discussed in section 7. 
\section{$SU(2)\times SU(2)$ chiral Lagrangian with excited
meson states }
In the usual $SU(2)\times SU(2)$ NJL model a local
(current--current) effective quark interaction is used
\be
L (\bar q, q) =
\int d^4 x \, \bar q (x) \left( i \partialslash - m^0 \right)
q (x) \; + \; L_{\rm int} ,
\label{L_NJL}
\ee
\be
L_{\rm int} &=& \sum_{a = 1}^3 \int d^4 x [ \frac{G_1}{2}
( j_{\sigma} (x) j_{\sigma} (x) +
j_{\pi}^a (x) j_{\pi}^a (x) )   \nonumber \\
&-& \frac{G_2}{2}( j_{\rho}^a (x) j_{\rho}^a (x)
 + j_{a_1}^a (x) j_{a_1}^a (x) )  ] ,
\label{L_int}
\ee
where $m^0$ is the current quark mass matrix. We suppose that
$m_u^0 \approx m_d^0 = m^0$. \\
$j_{\sigma , \pi , \rho , a_1} (x)$ denote the 
scalar, pseudoscalar, vector and axial-vector currents of the
quark fields, respectively,\footnote{The $\omega$-meson will be taken into
consideration at the end of the paper.}
\be
j_{\sigma} (x) &=& {\bar q}(x) q (x), \hspace{1.5cm}
j^a_{\pi} (x) = \bar q (x) i\gamma_5 \tau^a q (x), \nonumber \\
j^{a,\mu}_{\rho} (x) &=& \bar q (x) \gamma^{\mu} \tau^a q (x),
~~~~~~ j^{a,\mu}_{a_1} (x) = \bar q (x) \gamma_5 \gamma^{\mu}
\tau^a q (x).
\label{j_def}
\ee
Here $\tau^a$ are the Pauli matrices.
The model can be bosonized in the standard way by representing
the 4--fermion interaction as a Gaussian functional integral
over scalar, pseudoscalar, vector and axial-vector meson fields
\cite{volkov_83,volk_86,ebert_86}.
The effective meson Lagrangian, which is obtained by integration
over the quark fields, is expressed in terms of local meson
fields. By expanding the quark determinant in derivatives of the
local meson fields one then derives the chiral meson Lagrangian.
\par
The Lagrangian (\ref{L_int}) describes only ground--state
mesons. To include excited states, one has to introduce effective
quark interactions with a finite range.  In general, such
interactions require bilocal meson fields for bosonization
\cite{roberts_88,pervushin_90}. A possibility to avoid this
complication is the use of a separable interaction, which is
still of current--current form, eq. (\ref{L_int}), but allows for
non-local vertices (form factors) in the definition of the quark
currents, eqs. (\ref{j_def}),
\be
\tilde{L}_{\rm int} &=&
\int d^4 x \sum_{i = 1}^N \sum_{a = 1}^3 [ \frac{G_1}{2}
( j_{\sigma ,i} (x) j_{\sigma ,i} (x) +
j_{\pi , i}^a (x) j_{\pi , i}^a (x) ) \nonumber \\
 &-& \frac{G_2}{2} (j_{\rho , i}^a (x) j_{\rho , i}^a (x)
 + j_{a_1 , i}^a (x) j_{a_1 , i}^a (x) )] ,
\label{int_sep}
\ee
\be
j_{\sigma , i} (x) &=& \int d^4 x_1 \int d^4 x_2 \;
\bar q (x_1 ) F_{\sigma , i} (x; x_1, x_2 ) q (x_2 ),
\label{j_S} \\
j^a_{\pi , i} (x) &=& \int d^4 x_1 \int d^4 x_2 \;
\bar q (x_1 ) F^a_{\pi , i} (x; x_1, x_2 ) q (x_2 ),
\label{j_P} \\
j^{a,\mu}_{\rho , i} (x) &=& \int d^4 x_1 \int d^4 x_2 \;
\bar q (x_1 ) F^{a,\mu}_{\rho , i} (x; x_1, x_2 ) q (x_2 ).
\label{j_V} \\
j^{a,\mu}_{a_1 , i} (x) &=& \int d^4 x_1 \int d^4 x_2 \;
\bar q (x_1 ) F^{a,\mu}_{a_1 , i} (x; x_1, x_2 ) q (x_2 ).
\label{j_A}
\ee
Here, $F^{a,\mu}_{U, i}(x; x_1, x_2 )$, \,
$i = 1, \ldots N$, denote a set of non-local scalar,
pseudoscalar, vec\-tor and axial-vec\-tor quark ver\-tices
(in general momentum-- and spin--dependent),
which will be specified below. Upon bosonization
we obtain
\be
L_{\rm bos}(\bar q, q; \sigma, \pi, \rho, a_1) = \int d^4 x_1
\int d^4 x_2~ \bar q (x_1 ) [ ( i \partialslash_{x_2}
- m^0 ) \delta (x_1 - x_2 )      \nonumber \\
+ \int d^4 x  \sum_{i = 1}^N \sum_{a = 1}^3
( \sigma_i (x) F_{\sigma , i} (x; x_1, x_2 ) +
\pi_i^a (x) F_{\pi , i}^a (x; x_1, x_2) \nonumber \\
+ \rho_i^{a,\mu} (x) F_{\rho , i}^{a,\mu} (x; x_1, x_2)
+ a_{1,i}^{a,\mu} (x) F_{a_1 , i}^{a,\mu} (x; x_1, x_2) ) ]
q (x_2 ) \nonumber \\
- \int d^4 x \sum_{i = 1}^N \sum_{a = 1}^3
\left[ \frac{1}{2G_1} ( \sigma_i^2 (x) + \pi_i^{a\, 2} (x) )
- \frac{1}{2G_2} (\rho_i^{a,\mu\, 2} (x)+ a_{1,i}^{a,\mu\, 2} )
\right].
\label{L_sep}
\ee
This Lagrangian describes a system of local meson fields,
$\sigma_i (x)$, $\pi_i^a (x)$, $\rho^{a,\mu}_i (x)$,
$a_{1,i}^{a,\mu}$, $i = 1, \ldots N$, which interact with the
quarks through non-local vertices. These fields are not yet to be
associated with physical particles, which will be obtained after
determining the vacuum and diagonalizing the effective meson
Lagrangian.
\par
In order to describe the first radial excitations of mesons
(N = 2), we take the form factors in the form (see
\cite{volk_96} )
\be
F_{\sigma , 2} ({\bf k}) &=& f^{\pi} ({\bf k}),
\;\;\;\;\;\;\;\;\;\;\;\;
F^a_{\pi , 2} ({\bf k}) = i \gamma_5 \tau^a f^{\pi} ({\bf k}),
\nonumber  \\
F^{a,\mu}_{\rho , 2} ({\bf k}) &=& \gamma^\mu \tau^a f^{\rho}
({\bf k}),
\;\;\;\;
F^{a,\mu}_{a_1 , 2} ({\bf k}) = \gamma_5 \gamma^\mu \tau^a
f^{\rho} ({\bf k}),
\label{ffs}
\ee
\be
f^U ({\bf k}) = c^U ( 1 + d {\bf k}^2 ).
\label{ff}
\ee
We consider here the form factors in the momentum space and in
the rest frame of the mesons (${\bf P}_{meson}$ = 0; $k$ and
$P$ are the relative and total momentum of the quark-antiquark
pair, respectively). For the ground states of mesons one has
$f^{U,1} ({\bf k})$ = 1.
\par
After integrating over the quark fields in eq.(\ref{L_sep}),
one obtains the effective Lagrangian of the
$\sigma_1 , \sigma_2 , \pi_1^a,  \pi_2^a, \rho_1^{a,\mu}$,
$\rho_2^{a,\mu}$, $a_{1,1}^{a,\mu}$ and $a_{1,2}^{a,\mu}$,
fields.
\be
L(\sigma', \pi, \rho, a_1, \bar\sigma, \bar\pi, \bar\rho,
{\bar a}_1 ) =\;\;\;\;\;\;\;
~~~~~~~~~~ \nonumber \\
- \frac{1}{2 G_1} (\sigma^{'2} + \pi_a^2 + \bar\sigma^2 +
\bar\pi_a^2 )
+ \frac{1}{2 G_2} (\rho_a^2 + a_{1,a}^2 + \bar\rho_a^2
+ {\bar a}_{1,a}^2)
\nonumber \\
- i N_c \; {\rm Tr}\, \log [ i \partialslash - m^0 + \sigma'
+ (i \gamma_5  \pi_a +\gamma_\mu \rho^\mu_a + \gamma_5
\gamma_\mu a_{1,a}^\mu ) \tau^a  \nonumber \\
+ \bar\sigma f^{\pi} + (i \gamma_5  \bar\pi_a f^{\pi}
+\gamma_\mu \bar\rho^\mu_a f^{\rho} + \gamma_5 \gamma_\mu
{\bar a}_{1,a}^\mu f^{\rho} ) \tau^a ],
\label{12}
\ee
where we have put $\sigma_1 = \sigma', \sigma_2 = \bar\sigma, \pi_1 = \pi, \pi_2 = \bar\pi$ etc.
Now let us define the vacuum expectation of the $\sigma'$ field
\be
<\frac{\delta L}{\delta\sigma'}>_0 &=& - i N_c \; {\rm tr} \kint
\frac{1}{( \rlap/k - m^0 + <\sigma'>_0 )}
- \frac{<\sigma'>_0}{G_1} \; = \; 0 .
\label{gap_1}
\ee
Introduce the new sigma field which vacuum expectation is
equal to zero
\be
\sigma = \sigma' - <\sigma'>_0
\label{sigma}
\ee
and redefine the quark mass
\be
m = m^0 - <\sigma'>.
\label{m^0}
\ee
Then eq. (\ref{gap_1}) can be rewritten in the form of the usual
gap equation
\be
m = m^0 + 8 G_1 m I_1 (m),
\label{gap}
\ee
where
\be
I_n (m) = -i N_c \; \kint \frac{1}{(m^2 - k^2)^n}
\label{I_n}
\ee
and $m$ is the constituent quark mass.

\section{Effective Lagrangian for the ground and excited
states of the pions }
To describe the first excited states of pions and $\rho$-mesons,
it is necessary to use form factors $f^{\pi,\rho} ({\bf k})$
(see eq. (\ref{ff}))
\be
f^{\pi,\rho} ({\bf k}) = c^{\pi,\rho} ( 1 + d {\bf k}^2 ).
\label{ffq}
\ee
Following refs. \cite{volk_96, volk_97} we can fix
the slope parameter $d$ by using the condition
\be
I_1^f (m) = 0,
\label{I_1^f}
\ee
where
\be
I_1^{f..f} (m) = -i N_c \;
\kint \frac{f^U({\bf k})..f^U({\bf k})}{(m^2 - k^2)}.
\label{I_1^ff}
\ee
Eq. (\ref{I_1^f}) allows us to conserve the gap equation in
the form usual for the NJL model (see eq. (\ref{gap})), because the tadpole
with the excited scalar external field does not contribute to
the quark condensate and to the constituent quark mass.
\par
Using eq. (\ref{I_1^f}) and the values of $m$ and $\Lambda_3$ quoted in sec.5
we obtain for the 
slope parameter $d$ the value
\be
d = - 1.784~ GeV^{-2}.
\label{d_a}
\ee
\par
Now let us consider the free part of the Lagrangian (\ref{12}).
For the pions we obtain
\be
L^{(2)} (\pi) &=&
\half \sum_{i, j = 1}^{2} \sum_{a = 1}^{3}
\pi_i^a (P) K_{ij} (P) \pi_j^a (P) ,
\label{L_2}
\ee
where $K_{ij}(P)$ given by           
\be
K_{ij} (P) = -~\delta_{ij} \frac{1}{G_1} -
~i~ N_{\rm c} \; {\rm tr}\, \kint \left[
\frac{1}{\kslash + \half\Pslash - m}
i\gamma_5 f_i^{\pi}
\frac{1}{\kslash - \half\Pslash - m} i \gamma_5 f_j^{\pi}
\right]  , \nonumber \\
f_1^{\pi} \equiv 1, \hspace{2em} f_2^{\pi} \;\; \equiv \;\;
f^{\pi} ({\bf k}).~~~~~~~~~~
\label{K_full}
\ee
The integral (\ref{K_full}) is evaluated by expanding in the
meson field momentum, $P$. To order $P^2$, one obtains
\be
K_{11}(P) &=& Z_1 (P^2 - M_{\pi_1}^2 ),
\hspace{2em} K_{22}(P) \;\; = \;\; Z_2 (P^2 - M_{\pi_2}^2 ),
\nonumber \\
K_{12}(P) &=& K_{21}(P) \;\; = \;\;
\gamma P^2 ,
\label{K_matrix}
\ee
where
\be
Z_1 &=& 4 I_2 Z  , \hspace{2em} Z_2 \; = \; 4 I_2^{ff}
{\bar Z}, \hspace{2em} \gamma \; = \; 4 I_2^f Z,
\label{I_12}
\ee
\be
M_{\pi_1}^2 &=& (Z_1)^{-1}[\frac{1}{G_1}-8 I_1(m)]
= \frac{m^0}{4 m I_2 (m)}, \nonumber\\
M_{\pi_2}^2 &=& (Z_2)^{-1}[\frac{1}{G_1}-
8 I_1^{ff}(m)].
\label{Mp}
\ee
Here, $Z = 1 - \frac{6m^2}{M^2_{a_1}} \approx 0.7$,
${\bar Z} = 1 - {\Gamma^2_{\pi}}\frac{6m^2}{M^2_{a_1}} \approx
1$, $M_{a_1}$ is the mass of the $a_1$ meson and $\Gamma_{\pi}$ is
given below (see eq. (\ref{Gamma}) \footnote{The factors $Z$ and
$\bar Z$ appear when we take into account the transitions
$\pi_i \to a_1 \to \pi_j$.}.) $I_n,~I_n^f$ and $I_n^{ff}$ denote
the usual loop integrals arising in the momentum expansion
of the NJL quark determinant, but now with zero, one or two
factors $f^U ({\bf k})$, eqs.(\ref{ffq}), in the numerator (see
(\ref{I_1^ff}) and below )
\be
I_n^{f..f} (m) &=& -i N_{\rm c}
\kint \frac{f^U({\bf k})..f^U({\bf k})}{(m^2 - k^2)^n}.
\label{I_2^ff}
\ee
The evaluation of these integrals with a 3--momentum cutoff is
described {\em e.g.}\ in ref.\cite{ebert_93}. The integral over
$k_0$ is taken by contour integration, and the remaining
3--dimensional integral is regularized by the cutoff. Only the
divergent parts are kept; all finite parts are dropped. 
\par
After the renormalization of the pion fields
\be
\pi_i^{a r} = \sqrt{Z_i} \pi_i^a
\label{phi^r}
\ee
the Lagrangian (\ref{L_2}) takes the form
\be
L_\pi^{(2)} &=& \frac{1}{2} \left[ (P^2 - M^2_{\pi_1})~ \pi^2_1 +
2 \Gamma_\pi P^2~ \pi_1 \pi_2 + (P^2 - M^2_{\pi_2})~ \pi^2_2
\right].
\label{Lp}
\ee
Here
\be
\Gamma_{\pi} &=& \frac{\gamma}{\sqrt{Z_1 Z_2}} =
\frac{I_2^f\sqrt{Z}}{\sqrt{I_2 I_2^{ff}{\bar Z}}}.
\label{Gamma}
\ee
Using the additional transformation of the pion fields
\be
\pi^a = cos (\theta_{\pi} - \theta_{\pi}^0) \pi_1^{a r} -
cos (\theta_{\pi} + \theta_{\pi}^0) \pi_2^{a r}, \nonumber \\
\pi^{'a} = sin (\theta_{\pi} - \theta_{\pi}^0) \pi_1^{a r} -
sin (\theta_{\pi} + \theta_{\pi}^0) \pi_2^{a r},
\label{transf}
\ee
where
\be
sin \theta_{\pi}^0 = \sqrt{\frac{1 + \Gamma_{\pi}}{2}}, \quad
cos \theta_{\pi}^0 = \sqrt{\frac{1 - \Gamma_{\pi}}{2}}
\label{theta_ch}
\ee
the Lagrangian (\ref{Lp}) takes the diagonal form
\be
L_\pi^{(2)} &=& \half (P^2 - M_\pi^2)~ \pi^2 +
\half (P^2 - M_{\pi'}^2)~ \pi^{' 2}.
\label{L_pK}
\ee
Here
\be
M^2_{\pi, \pi'} = \frac{1}{2 (1 - \Gamma^2_\pi)}
[M^2_{\pi_1} + M^2_{\pi_2}~
\pm~ \sqrt{(M^2_{\pi_1} - M^2_{\pi_2})^2 +
(2 M_{\pi_1} M_{\pi_2} \Gamma_\pi)^2}]
\label{MpK}
\ee
and
\be
\tan 2 \bar\theta_{\pi} = \sqrt{\frac{1}{\Gamma^2_{\pi}} -1}~
\left[ \frac{M^2_{\pi_1} - M^2_{\pi_2}}{M^2_{\pi_1} +
M^2_{\pi_2}} \right] =
- \tan 2 \bar\theta_\pi^0~
\left[ \frac{M^2_{\pi_1} - M^2_{\pi_2}}{M^2_{\pi_1} +
M^2_{\pi_2}} \right],~~(2\theta_{\pi} = 2{\bar \theta_{\pi}}+ \pi) .
\label{tan}
\ee
In the chiral limit $M_{\pi_1} \to 0$,
$\theta_\pi \to \theta_\pi^0$ (see eqs. (\ref{Mp}, \ref{tan}) )
we obtain
\be
M_\pi^2 &=& M_{\pi_1}^2 \; + \; {\cal O}(M_{\pi_1}^4 ),
\label{Mp_ch}
\ee
\be
M_{\pi'}^2 &=& \frac{M_{\pi_2}^2 + M_{\pi_1}^2 \Gamma_\pi}
{1 - \Gamma^2_\pi} \; + \; {\cal O}(M_{\pi_1}^4 ).
\label{Mp'_ch}
\ee
Thus, in the chiral limit the effective Lagrangian
eq. (\ref{Lp}) describes a massless Goldstone pion,
$\pi$, and a massive particle, $\pi'$.
\par
For the weak decay constants of the pions we obtain
(see \cite{volk_96})
\be
F_{\pi} &=& 2 m \sqrt{Z I_2(m)}~ cos (\theta_\pi
-\theta_\pi^0),  \nonumber \\
F_{\pi'} &=& 2 m \sqrt{Z I_2(m)}~sin (\theta_\pi
-\theta_\pi^0).
\label{f_p}
\ee
In the chiral limit we have 
\be
F_\pi = \frac{m}{g_\pi},~~~~F_{\pi'} = 0.
\label{f_ch}
\ee
with $g_{\pi} = Z_1^{-1/2}$ which is just the Goldberger-Treimann relation
for the coupling
constant $g_\pi$. The matrix elements of the divergence of
the axial current between meson states and the vacuum equal
(PCAC relations)
\be
\langle 0 | \partial^\mu A_\mu^a | \pi^b \rangle &=&
M_\pi^2 F_\pi \delta^{ab} ,
\label{A_phi} \\
\langle 0 | \partial^\mu A_\mu^a | \pi^{\prime\, b} \rangle &=&
M_{\pi'}^2 F_{\pi'} \delta^{ab}
\label{A_phi'} .
\ee
Then from eqs. (\ref{Mp_ch}) and (\ref{f_ch}) we can see that
the axial current is conserved in the chiral limit, because
its divergence equals zero, according to the low-energy
theorems.
\par
\section{Effective Lagrangian for ground and excited
states of the $\rho$($\omega$)-mesons}
The free part of the effective Lagrangian (\ref{12}) describing
the ground and excited states of the $\rho$- and $\omega$-mesons has the form
\be
L^{(2)} (\rho,\omega) &=&
- \half \sum_{i, j = 1}^{2} \sum_{a = 0}^{3} \rho_i^{\mu a} (P)
R_{ij}^{\mu\nu} (P) \rho_j^{\nu a}(P) ,
\label{LV_2}
\ee
where 
\be
\sum_{a = 0}^{3} (\rho_i^{\mu a})^2 = (\omega_i^{\mu})^2 + 
(\rho_i^{0 \mu})^2 + 2 \rho_i^{+ \mu} \rho_i^{- \mu} ~~~
\label{V^a}
\ee
and
\be
R_{ij}^{\mu \nu} (P) =
\frac{\delta_{ij}}{G_2} g^{\mu\nu}
- i N_{\rm c} \; {\rm tr}\, \kint \left[
\frac{1}{\kslash + \half\Pslash - m}\gamma^\mu f_i^{\rho}
\frac{1}{\kslash - \half\Pslash - m}\gamma^\nu f_j^{\rho}
\right]  , \nonumber \\
f_1^{\rho} \equiv 1, \hspace{2em} f_2^{\rho} \;\; \equiv \;\;
f^{\rho} ({\bf k}).\hspace{3cm}
\label{R_full}
\ee
To order $P^2$, one obtains
\be
R_{11}^{\mu\nu} &=& W_1 [P^2 g^{\mu\nu} - P^\mu P^\nu -
g^{\mu\nu} M_{\rho_1}^2], \nonumber \\
R_{22}^{\mu\nu} &=& W_2 [P^2 g^{\mu\nu} - P^\mu P^\nu -
g^{\mu\nu} M_{\rho_2}^2], \nonumber \\
R_{12}^{\mu\nu} &=& R_{21}^{\mu\nu} = \gamma_\rho
[P^2 g^{\mu\nu} - P^\mu P^\nu ].
\label{R_ij}
\ee
Here
\be
W_1 &=& \frac{8}{3} I_2,~~~W_2 = \frac{8}{3} I_2^{ff},~~~
\gamma_\rho = \frac{8}{3} I_2^{f}, \\
M_{\rho_1}^2 &=& (W_1 G_2)^{-1} , ~~~~
M_{\rho_2}^2 = (W_2 G_2)^{-1} .
\label{WM}
\ee
After renormalization of the $\rho$($\omega$)-meson fields
\be
\rho_i^{\mu a r} = \sqrt{W_i}~\rho_i^{\mu a}
\label{V^r}
\ee
we obtain the Lagrangian
\be
L_\rho^{(2)} &=& - \half [( g^{\mu\nu} P^2 - P^\mu P^\nu -
g^{\mu\nu} M^2_{\rho_1}) \rho^\mu_1 \rho^\nu_1 \nonumber \\
&+& 2 \Gamma_\rho  ( g^{\mu\nu} P^2 - P^\mu P^\nu) \rho_1^\mu
\rho_2^\nu + ( g^{\mu\nu} P^2 - P^\mu P^\nu -
g^{\mu\nu} M^2_{\rho_2}) \rho^\mu_2 \rho^\nu_2 ],
\label{L2_V1}
\ee
where
\be
\Gamma_{\rho} = \frac{I_2^{f}(m)}
{\sqrt{I_2(m)I_2^{ff}(m)}}.
\label{GammaV}
\ee
By trasforming the $\rho$-meson fields similarly to eqs.
(\ref{transf}) used for pions, the Lagrangian
(\ref{L2_V1}) takes the diagonal form
\be
L^{(2)}_{\rho, \rho'} = - \half \left[ (g^{\mu\nu} P^2 -
P^\mu P^\nu - M^2_{\rho}) \rho^{\mu} \rho^{\nu} 
+ (g^{\mu\nu} P^2 - P^\mu P^\nu -
M^2_{\rho'} ) \rho^{' \mu} \rho^{' \nu} \right],
\label{LDV}
\ee
where $\rho$ and $\rho'$ are the physical ground and
excited $\rho$-meson states and
\be
M^2_{\rho, \rho'} = \frac{1}{2(1 - \Gamma^2_\rho)}~
\left[M^2_{\rho_1} + M^2_{\rho_2}~ \pm~ \sqrt{(M^2_{\rho_1}-
M^2_{\rho_2})^2 + (2 M_{\rho_1}M_{\rho_2} \Gamma_\rho)^2}\right] .
\label{Mrho}
\ee
\par
The same formulae are valid for the $\omega$-meson.
\par
\section{Numerical estimates}
We can now estimate numerically the masses of the pions
and $\rho$-mesons and the weak decay constants $F_\pi$ and
$F_{\pi'}$ in our model.
\par
Because the mass formulae and others equations ( for instance,
Goldberger -- Treimann relation etc. ) have new forms in
the extended NJL model with excited states of mesons as compared with
the usual NJL model, the values of basic parameters ($m$,
$\Lambda_3$, $G_1$, $G_2$) could change. However, this does not happen for
the parameters $m=280~{\rm MeV}$,
$\Lambda_3 = 1.03~{\rm GeV}~$ and
$G_1 = 3.47~{\rm GeV}^{-2}$ (see ref. \cite{ebert_93}), because the condition
(\ref{I_1^f})
conserves the gap equation in the old form 
and one can satisfactorily describe the weak decay constant $F_\pi$ and
the decay $\rho \to 2\pi$ in the extended  model,too, using $m=280~{\rm MeV}$
and the cutoff parameter $\Lambda_3 = 1.03~{\rm GeV}$ (see below).
$G_1$ does not change in the extented model, because $M_\pi \approx 
M_{\pi_1}$. However, for the coupling constant $G_2$ the new value $G_2
=12.5~{\rm GeV}^{-2}$ will be used, which differs
from the former value $G_2 = 16~{\rm GeV}^{-2}$ (see ref.
\cite{ebert_93}). It is a consequence of the fact that the mass
$M_{\rho_1}$ noticeably differs from the physical mass $M_\rho$
of the ground state $\rho$.
\par
Using these basic parameters, the slope
parameter $d = -1.784~{\rm GeV}^{-2}$ (see eq. (\ref{d_a}))
and choosing the form factor parameters
$c^{\pi} = 1.37$ and $c^{\rho} = 1.26$, one finds
\be
M_\rho = 768.3~{\rm MeV},~~ M_{\rho'} = 1.49~{\rm GeV},~~
M_\pi = 136~{\rm MeV},~~M_{\pi'} = 1.3~{\rm GeV}.
\label{Mprot}
\ee
The experimental values are
\be
M^{exp}_\rho &=& 768.5 \pm 0.6~{\rm MeV},~~~ M^{exp}_{\rho'} =
1465 \pm 25~{\rm MeV}, \nonumber \\
M^{exp}_{\pi^+} &=& 139.57~{\rm MeV},~~~~~~~~~M^{exp}_{\pi^0} = 134.98~
{\rm MeV}, \nonumber \\
M^{exp}_{\pi'} &=& 1300 \pm 100~{\rm MeV}.
\label{Mproe}
\ee
\be
F_\pi = 93~{\rm MeV},~~~~ F_{\pi'} = 0.57~{\rm MeV}, \nonumber \\
\frac{F_{\pi'}}{F_{\pi}} \approx
- \cot 2 \theta_\pi^0~\left( \frac{M_\pi}{M_{\pi'}} \right)^2 \approx
0.5 (\frac{M_{\pi}}{M_{\pi'}})^2.
\label{ff'}
\ee
\section{Decays $\rho \to 2 \pi, \pi' \to \rho \pi, \rho' \to 2\pi, 
\rho' \to \omega \pi$ and $\omega' \to \rho \pi$.}
\par
Let us show how the decay widths of the ground and excited states
of mesons are calculated in our model. For that we start with the
decay $\rho \to 2\pi$. The amplitude describing this decay has
the form
\be
T_{\rho \to 2\pi} = i~\frac{g_\rho}{2}~\epsilon_{ijk}~(p_j -
p_k)^\nu~\rho^i_\nu \pi^j \pi^k,
\ee
where $p_{j,k}$ are the pion momenta and $\epsilon_{ijk}$ is the
antisymmetric tensor. Using the value $\alpha_{\rho} = 
\frac{g^2_{\rho}}{4 \pi} \approx 3~~(g_\rho \approx 6.1)$ of refs. 
\cite{volkov_83,volk_86,ebert_86} we obtain for the decay width 
\be
\Gamma_{\rho \to 2\pi} =  \frac{\alpha_\rho}{12~M_\rho^2}~
(M_\rho^2 - 4~M_\pi^2)^{3/2} = 151.5~{\rm MeV}.
\ee
The experimental value is \cite{Rev_96}
\be
\Gamma_{\rho \to 2\pi} = 150.7 \pm 1.2~{\rm MeV} 
\ee
\par
Now let us calculate this amplitude in our model with the excited
states of mesons. For that we rewrite the amplitude $T_{\rho \to
2\pi}$ in the form
\be
T_{\rho \to 2\pi} = i~c_{\rho \to 2\pi}~\epsilon_{ijk}~(p_j -
p_k)^\nu~\rho^i_\nu \pi^j \pi^k,
\ee
and calculate the factor $c_{\rho \to 2\pi}$ in the new model.
Using eqs. (\ref{phi^r}), (\ref{transf}) and (\ref{V^r}) we
can find the following expressions for the meson fields $\pi_i$
and $\rho_i$ from the Lagrangian (\ref{12}) expressed in terms of the 
physical states $\pi, \pi'$ and $\rho, \rho'$ ($\alpha = \theta_{\pi}, 
\beta = \theta_{\rho}$)
\be
\pi_1 =\frac{ sin(\alpha+\alpha_0) \pi - cos(\alpha+\alpha_0)
\pi'}{\sqrt{Z_1} sin2\alpha_0}, \nonumber \\
\pi_2 =\frac{ sin(\alpha-\alpha_0) \pi - cos(\alpha-\alpha_0)
\pi'}{\sqrt{Z_2} sin2\alpha_0},
\label{pi}
\ee
\be
\rho_1 = \frac{sin(\beta+\beta_0) \rho - cos(\beta+\beta_0)\rho'}
{sin2\beta_0 \sqrt{8/3~I_2}}, \nonumber \\
\rho_2 = \frac{sin(\beta-\beta_0) \rho - cos(\beta-\beta_0)\rho'}
{sin2\beta_0 \sqrt{8/3~I_{2,\rho}^{ff}}},
\label{ro}
\ee
or, using the values $I_2 = 0.04, I_{2,\rho}^{ff} = 0.0244$,
$\alpha = 59.5^o,~\alpha_0 = 59.15^o,~\beta = 79.9^o,~
\beta_0 = 61.5^o$, 
we obtain \footnote{Analogous formulae are obtained for the $\omega$-meson.}
\be
\pi_1 = \frac{0.878 \pi +0.48 \pi'}{0.88 \sqrt{Z_1}},~~~
\pi_2 = \frac{0.0061 \pi - \pi'}{0.88 \sqrt{Z_2}}, \nonumber \\
\rho_1 = (0.744 \rho + 0.931 \rho')~g_\rho/2,~~~
\rho_2 = (0.48~ \rho - 1.445~\rho')~g_\rho/2.
\label{piro}
\ee
The decay $\rho \to 2 \pi$ is described by the quark triangle
diagrams with the vertices \\
$\rho_1 (\pi^2_1 + 2\pi_1\pi_2 + \pi_2^2)$ and
$\rho_2 (\pi^2_1 + 2\pi_1\pi_2 + \pi_2^2)$ (see Fig.1).
Using eqs. (\ref{pi}), (\ref{ro}) and (\ref{piro}) the factor 
$c_{\rho \to 2\pi}$ is given by 
\footnote{Taking into account the $\pi \to a_1$ transitions on
the external pion lines we obtain additional factors $Z$
($\bar{Z}$) in the numerators of our triangle diagrams, which
cancel corresponding factors in $Z_i$ (see eqs. (\ref{I_12}),
(\ref{pi}) and ref. \cite{volk_86}). Therefore, in future
we shall ignore the factors $Z$ ($\bar{Z}$) in $Z_i$. }
\be
c_{\rho \to 2\pi} = c_{\rho_1 \to 2\pi} + c_{\rho_2 \to 2\pi} =
0.975~g_\rho/2,
\ee
\be
c_{\rho_1 \to 2\pi} &=& \frac{sin(\beta + \beta_0)}{sin^2
2\alpha_0~sin 2\beta_0~\sqrt{8/3~I_2}}~[(sin(\alpha +
\alpha_0))^2 + 2 sin(\alpha + \alpha_0) sin(\alpha - \alpha_0)
\Gamma_\pi   \nonumber  \\
&+& (sin(\alpha - \alpha_0))^2 = sin^2 2\alpha_0] =
\frac{sin(\beta + \beta_0)}{sin 2\beta_0~\sqrt{8/3~I_2}}
= 0.745~g_\rho/2,  \nonumber \\
c_{\rho_2 \to 2\pi} &=& \frac{sin(\beta - \beta_0)}{sin^2
2\alpha_0~ sin 2\beta_0~\sqrt{8/3~I_{2,\rho}^{ff}}}~
[(sin(\alpha + \alpha_0))^2~\frac{I_2^f}{I_2}  \nonumber \\
&+& 2 sin(\alpha + \alpha_0) sin(\alpha -
\alpha_0) \frac{I_2^{ff}}{\sqrt{I_2~I_2^{ff}}} +
(sin(\alpha - \alpha_0))^2 \frac{I_2^{fff}}{I_2^{ff}}] = 0.227~
g_\rho/2.
\label{cro}
\ee
Here we used the values
$I_2 = 0.04,~I_2^f = 0.0185,~I_2^{ff} = 0.0289,~I_2^{fff} =
0.0224$ and the relation $\Gamma_\pi = - cos 2\alpha_0$ (see eqs.
(\ref{theta_ch})). Then the decay width $\rho \to 2 \pi$ is equal to
\be
\Gamma_{\rho \to 2\pi} = 149~{\rm MeV}.
\ee
In the limit $f = 0$ ($\alpha = \alpha_0, \beta = \beta_0$) from
eqs. (\ref{cro}) one finds
\be
c_{\rho \to 2\pi} = c_{\rho_1 \to 2\pi} = g_\rho/2,~~~
c_{\rho_2 \to 2\pi} = 0.
\ee
\par
Now let us consider the decay $\pi' \to \rho \pi$. The amplitude
of this decay has the form
\be
T_{\pi' \to \rho \pi} = i~c_{\pi' \to \rho \pi}~\epsilon_{ijk}~
(p_j + p_k)^\nu~\rho^i_\nu \pi^j \pi^k,
\ee
where
\be
c_{\pi' \to \rho \pi} = c_{\pi' \to \rho_1 \pi} + c_{\pi' \to
\rho_2 \pi}.
\ee
Then for $c_{\pi' \to \rho_1 \pi}$ we obtain
\be
c_{\pi' \to \rho_1 \pi} &=& \frac{2}{(sin 2\alpha_0)^2}~
[-sin(\alpha+\alpha_0) cos(\alpha+\alpha_0) - sin 2\alpha~
\Gamma_\pi - sin(\alpha-\alpha_0)
cos(\alpha-\alpha_0)  \nonumber  \\
&=& - sin 2\alpha cos 2\alpha_0 + sin 2\alpha cos 2\alpha_0 = 0]~
\frac{sin(\beta+\beta_0)}{sin 2\beta_0}~g_\rho/2 = 0,
\label{cro1}
\ee
\be
c_{\pi' \to \rho_2 \pi} = \frac{2}{(sin 2\alpha_0)^2}~
[-sin(\alpha+\alpha_0) cos(\alpha+\alpha_0) \frac{I_2^f}{I_2} -
sin 2\alpha \frac{I_2^{ff}}{\sqrt{I_2~I_2^{ff}}} \nonumber \\
- sin(\alpha-\alpha_0) cos(\alpha-\alpha_0)
\frac{I_2^{fff}}{I_2^{ff}}]~
\frac{sin(\beta - \beta_0)}{sin 2\beta_0}~
\sqrt{\frac{I_2}{I_2^{ff}}}~g_\rho/2 = - 0.573~g_\rho/2.
\label{cro2}
\ee
For the decay width $\pi' \to \rho \pi$ we obtain
\be
\Gamma_{\pi' \rightarrow \rho \pi} &=& \frac{c_{\pi' \to \rho
\pi}^2}{4\pi M^3_{\pi'}M^2_{\rho}} \left[ M^4_{\pi'} + M^4_{\rho}
+ M^4_{\pi} - 2(M^2_{\pi'}M^2_{\rho} + M^2_{\pi'}M^2_{\pi} +
M^2_{\rho}M^2_{\pi} )  \right]^{3/2} \nonumber  \\
&=& 220~{\rm MeV}.
\label{Gpi'1}
\ee
This value is in agreement with the
experimental data \cite{Rev_96}
\be
\Gamma^{total}_{\pi'} = 200 - 600~ {\rm MeV}.
\ee
\par
The decay $\pi' \to \sigma \pi$ in our model gives only a small
contribution to the total decay width of $\pi'$.
\par
 For the decay $\rho' \to 2\pi$
we obtain in our model the result
\be
\Gamma_{\rho' \to 2\pi} \approx 22~{\rm MeV}.
\ee
All our results are in agreement with the results of a relativized
potential quark model with the $3P_0$-mechanism of meson decays \cite{gov}.
\par
 In conclusion of this section let us calculate the decay widths of the 
processes $\rho' \to \omega \pi$ and $\omega' \to \rho \pi$.  
These decays go through anomalous triangle quark loop diagrams.
The amplitude of the decay $\rho' \to \omega \pi$ takes the form
\be
T^{\mu \nu}_{\rho' \to \omega \pi} = \frac{3 \alpha_{\rho} 
c_{\rho' \to \omega \pi}}{2 \pi F_{\pi}}~\epsilon
_{\mu \nu \rho \sigma}~
q^{\rho} p^{\sigma},
\ee
where $q$ and $p$ are the momentum of the $\omega$ and $\rho'$ meson,
respectively. The factor $c_{\rho' \to \omega \pi}$ is similar to the
factors $c_{\rho \to 2\pi}$ and $c_{\pi' \to \rho \pi}$ in previous
equations and arises from the four triangle quark diagrams with vertices 
$\pi_1(\rho_1 \omega_1 + \rho_2 \omega_1 + \rho_1 \omega_2 + 
\rho_2 \omega_2)$
\footnote{We shall neglect the diagrams with vertices $\pi_2$, because their
contribution to the ground state of the pion is very small (see 
eq.(\ref{piro})).}. Using the estimate
\be
c_{\rho' \to \omega \pi} \approx - 0.3,
\ee
we obtain for the decay width
\be
\Gamma_{\rho' \rightarrow \omega \pi} &=& \frac{3}{2 \pi M^3_{\rho'}}~
(\frac{\alpha_{\rho}~c_{\rho' \to \omega \pi}}{8~\pi~F_{\pi} })^2~
\left[ M^4_{\rho'} + M^4_{\omega}
+ M^4_{\pi} - 2(M^2_{\omega}M^2_{\rho'} + M^2_{\rho'}M^2_{\pi} +
M^2_{\omega}M^2_{\pi} )  \right]^{3/2} \nonumber  \\
&\approx& 75~{\rm MeV}.
\ee
For the decay $\omega' \to \rho \pi$ we have the relation
\be
\Gamma_{\omega' \to \rho \pi} \approx 3~\Gamma_{\rho' \to \omega \pi}
\ee
leading to the estimate
\be
\Gamma_{\omega' \to \rho \pi} \approx 225~{\rm MeV}. 
\ee
The experimental values are \cite{clegg}
\be
\Gamma^{exp}_{\rho' \to \omega \pi} =  
0.21~\Gamma^{tot}_{\rho'} = 65.1~\pm~12.6~{\rm MeV} 
\ee
and \cite{Rev_96}
\be
\Gamma^{exp}_{\omega' \to \rho \pi} = 174~\pm~60~{\rm MeV}. 
\ee
Finally, let us quote the ratio of the decay widths $\rho' \to \omega \pi$
and $\rho' \to 2\pi$
\be
\frac{\Gamma_{\rho' \to 2 \pi}}{\Gamma_{\rho' \to \omega \pi} } \approx 0.3,
\ee
which has to be compared with the experimental value 0.32 (see \cite{clegg}).
\par
Thus, we can see that all our estimates are in satisfactory agreement with 
experimental data.
\par
\section{Summary and conclusions}
 Our calculations have shown that the main decay of the $\rho$-meson,
$\rho \to 2\pi$, changes very weakly after including the excited meson
states into the NJL model. The main part of this decay (75\%) comes from
the $\rho$-vertex without form factor, whereas the remaining 25\% of the 
decay are due to the $\rho$-vertex with form factor. As a result,  the new
coupling constant $g_{\rho}$ turns out to be very close to the former value.
\par 
For the decay $\pi' \to \rho \pi$ we have the opposite situation.
Here the channel connected with the $\rho$-vertex without form factor
is closed, because the states $\pi$ and $\pi'$ are orthogonal to
each other, and the total decay width of
$\pi' \to \rho \pi$ is defined by the channel going through the $\rho$-
vertex with form factor. As a result, we obtain 
the quoted  value which 
satisfies the experimental data \cite{Rev_96}.
The decay $\pi' \to \sigma \pi$ gives only very small corrections to 
the total decay width of $\pi'$. Notice that these results are in agreement
with the results obtained in the relativized version of the $3P_1$ potential
model \cite{gov}.
\par
For the decay $\rho' \to 2\pi$ we obtain a strong compensation
of the contributions of the two channels, connected with $\rho$-vertices
with and without form factors, and the corresponding  decay width is equal to
22 MeV . Again this value is very close to the result of ref.\cite{gov}.
\par
It should be emphasized, that the decays $\rho' \to \omega \pi$ 
and $\omega' \to 
\rho \pi$ belonging to quite another class of quark loop diagrams
(``anomaly diagrams'') 
are also satisfactorily described by our model. In future applications we
are planning to describe the decays of other pseudoscalar, scalar,
vector and axial-vector mesons of the U(3) flavour group, too.
\par 
Finally, it is worth mentioning that concerning in particular the bosonization
program of QCD, the description of excited mesons has constantly attracted
considerable interest by many authors (see e.g. \cite{andr, efim_96}
and references therein).

{\large \bf Acknowledgments}

The authors would like to thank Dr.S.B.Gerasimov for fruitful
discussions. M.N. was supported in part by the Slovak Scientific
Grant Agency, Grant VEGA No. 2/4111/97. M.K.V. was partly supported
by the Heisenberg-Landau Grant and
the Graduiertenkolleg of the Humboldt University, Berlin.

\newpage

\begin{figure}[!htb]
  \centering
  \epsfig{file=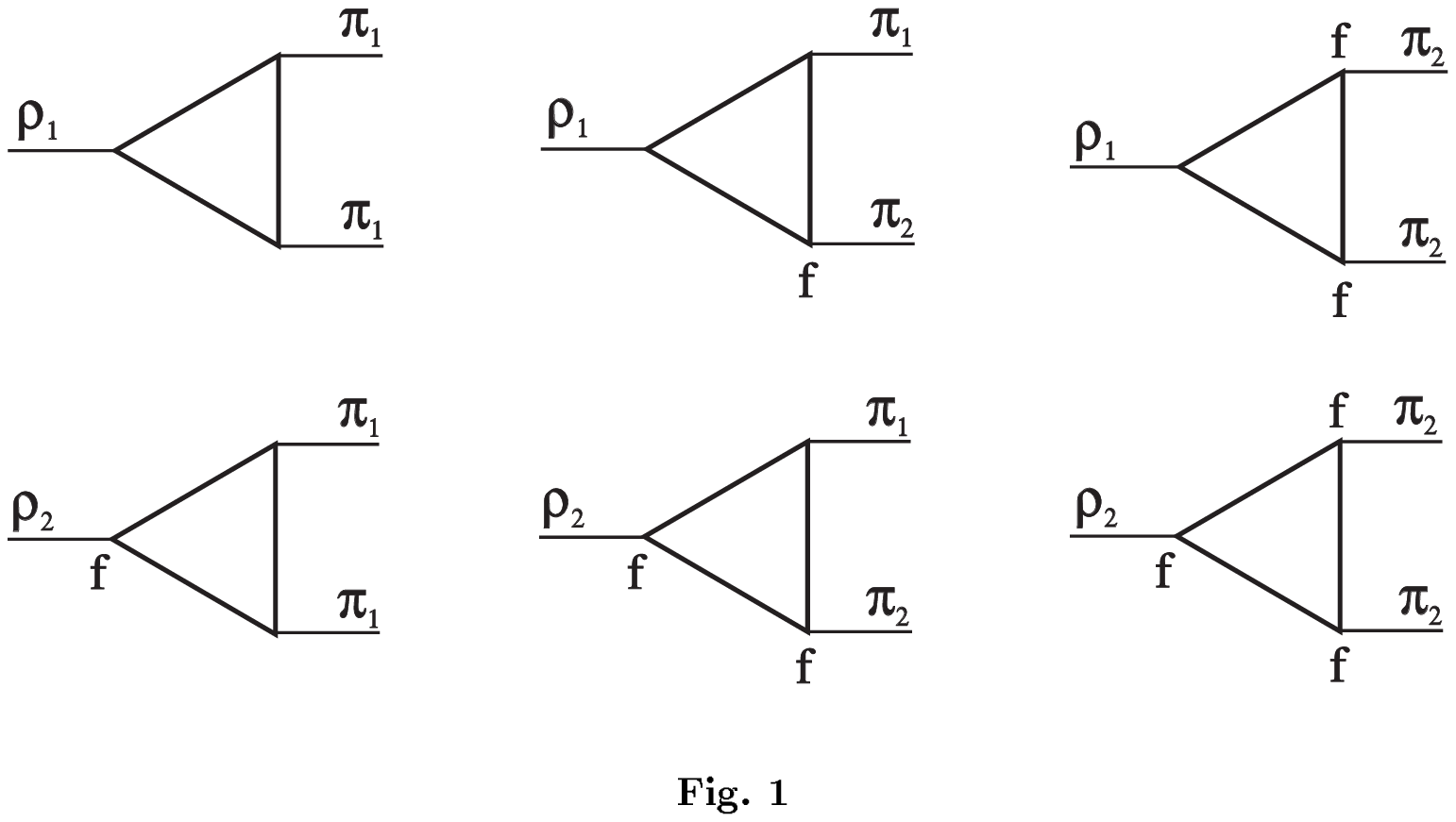}
\end{figure}

\begin{center}
\section*{\bf Figure caption}
\end{center}
\vskip1.0truecm
\begin{description}
\item{\bf Fig.1.}~Triangle diagrams describing the decays:
\item{~~~~~~}a) $\rho \to 2\pi$, when we consider the parts of
$\rho_i$ and $\pi_i$ meson fields corresponding only to the 
ground states $\rho$ and $\pi$;
\item{~~~~~~}b) $\pi'\to \rho\pi$, when we consider the parts
of the $\rho_i$ corresponding to the ground state $\rho$ and take
for one of the pions $\pi_i$ the part corresponding to the ground
state $\pi$ and for the other the excited pion state $\pi'$.
\item{~~~~~~}c) $\rho' \to 2\pi$, when we consider the parts of $\rho_i$
corresponding to the excited $\rho'$ and take for the $\pi_i$ only the
ground state pions.
\end{description}


\begin{thebibliography}{99}
%
\bibitem{Rev_96} Review of Particle Properties,
Phys.\ Rev.\ {\bf D 54} 1 (1996).
%
\bibitem{volk_96}
M.K.\ Volkov and Ch.\ Weiss, Bochum Univ. preprint
RUB--TPII--12/96 (1996);
hep--ph/9608347; to be published in Phys. Rev. D (1997).
%
\bibitem{volk_97}
M.K.\ Volkov, JINR preprint E2-96-482, Dubna, 1996; Yad. Fiz.
{\bf 60} N 10 (1997).
%
\bibitem{feynman_71} R.P.\ Feynman, M.\ Kislinger and F.\
Ravndal, Phys.\ Rev.\ {\bf D 3} 2706 (1971).
%
\bibitem{volkov_83}
D. Ebert and M.K. Volkov, Z.\ Phys.\ {\bf C 16} 205 (1983);\\
M.K. Volkov, Ann.\ Phys.\ (N.Y.) {\bf 157} 282 (1984).
%
\bibitem{volk_86}
M.K. Volkov, Sov.\ J.\ Part.\ Nucl.\  {\bf 17} 186 (1986).
%
\bibitem{ebert_86} D. Ebert and H. Reinhardt, Nucl.\ Phys.\
{\bf B 271} 188 (1986).
%
\bibitem{roberts_88}
C.D.\ Roberts, R.T.\ Cahill and J.\ Praschifka, Ann.\ Phys.\
(N.Y.) {\bf 188} 20 (1988).
%
\bibitem{pervushin_90} D.\ Ebert, V.N.\ Pervushin, H.\ Reinhardt, 
Sov.\ Part.\ Nucl. {\bf 10} 444 (1979);  \\
V.N.\ Pervushin {\em et al.},
Fortschr.\ Phys.\ {\bf 38} 333 (1990); \\
Yu.L.\ Kalinovsky {\em et al.}, Few--Body Systems
{\bf 10} 87 (1991).
%
\bibitem{ebert_93} D.\ Ebert, Yu.L.\ Kalinovsky, L. M\"unchow and
M.K.\ Volkov, Int.\ J.\ Mod.\ Phys. {\bf A 8} 1295 (1993).
%
\bibitem{gov} S.B.\ Gerasimov, A.B.\ Govorkov, Z.\ Phys. {\bf C 13}
43 (1982).
%
\bibitem{clegg} A.B.\ Clegg, A.\ Donnachie, Z.\ Phys. {\bf C 62} 455 (1994). 
%
\bibitem{andr} A.A.\ Andrianov and V.A.\ Andrianov,
Int.\ J.\ Mod.\ Phys.\ {\bf A 8} 1981 (1993);
Nucl.\ Phys.\ {\bf B} (Proc.\ Suppl.) {39 B, C} (1993).
%
\bibitem{efim_96} Ja.V.\ Burdanov, G.V.\ Efimov, S.N.\ Nedelko, S.A.\ Solunin,  Phys.\ Rev. {\bf D 54} 4483 (1996).
\end{thebibliography}
\end{document}